# Conceptual Foundations of Special and General Relativity

**Charles Francis**

**Abstract:**
   The *k*-calculus was advocated by Hermann Bondi as a means of explaining special relativity using only GCSE level mathematics and ideas. We review the central derivations, using proofs which are only a little more elegant than those in Bondi's books, and extend his development to include the scalar product and the mass shell condition. As used by Bondi, *k* is the Doppler red shift, and we extend the *k* calculus to include the gravitational red shift and give a derivation of Newton's law of gravity using only first year calculus and basic quantum mechanics.

Charles Francis
Clef Digital Systems Ltd.
Lluest, Neuaddlwyd
Lampeter
Ceredigion
SA48 7RG

28/9/99

# Conceptual Foundations of Special and General Relativity

## 1  The Fabric of Space-time

Riemann's mathematical definition [1] of a manifold deliberately ignores the question of whether the manifold represents something real. Riemann had this to say (Clifford's translation, quoted in [2])

Either, therefore, the reality which underlies space must form a discrete manifold, or we must seek the ground of its metric relations outside it.

The purpose of the present paper is to show that the laws of special and general relativity can derived from a straightforward treatment of measurement which does not require either the assumption of a manifold, or an understanding of the tensor calculus. Instead the *k*-calculus, which was popularised by Bondi [3] for special relativity, is used and extended to characterise the non-Euclidean geometry of space-time. In the *k*-calculus for special relativity, *k* is equivalent to Doppler shift. In the treatment of general realtivity *k* is gravitational red shift, and fully characterises the geometry of a local region of space, without resource to the tensor calculus. We illustrate the power of the method with a derivation of Newton's law of gravity from basic quantum mechanics and simple calculus.

The conceptual basis of the treatment is that quantum electrodynamics has shown that the exchange of photons is responsible for the electromagnetic force, and so for all the structures of matter in our macroscopic environment. But as Bondi pointed out *'with our modern outlook and modern technology the Michelson-Morley experiment is a mere tautology"* [4]. It is not unreasonable, therefore, to postulate that photon exchange generates all the geometrical relationships in the macroscopic environment, just as it generates these relationships in the results of measurement by radar.

## 2  Co-ordinate Systems

There is room for confusion between two very similar questions, 'What is time?' and 'What is the time?'. The first question has something to do with consciousness, and our perception of time as a flow from past to future. It admits no easy answer, but is quite distinct from the second question and only the second question is relevant in the definition of space-time co-ordinates. The answer to the question 'What is the time?' is always something like 4:30 or 6:25.

**Definition:** The time is a number read from a clock.

There are many different types of clock, but every clock has two common elements, a repeating process and a counter. The rest of the mechanism converts the number of repetitions to conventional units of time. A good clock should provide accurate measurement and it should give a uniform measure of time. We cannot count less than one repetition of the process in the clock, so for accurate measurement the process must repeat as rapidly as possible. In a uniform clock, the repeating process must repeat each time identical to the last, uninfluenced by external matter.

A clock defines the time, but does so only at one place. A space-time co-ordinate system also requires a definition of distance, and a definition of time at a distance from the clock. This is provided for by the radar method (in practice the frequency of radar is irrelevant, and the definition refers to light of any frequency).

**Definition:** The distance of an event is half the lapsed time for light to go from the clock to the event and return to the clock. the time at which the signal is reflected is the mean time between when it is sent and when it returns.

The radar method defines distance in units of time, and this paper will use natural units in which the speed of light is 1, and speed is measured as a fraction of the speed of light. To restore conventional units substitute $v \rightarrow v/c$ Radar is preferred to a ruler, because it applies directly to both large and small distances, and because a single measurement can be used for both time and space co-ordinates. The radar method also measures direction and it will be seen that the algebra is formally identical for 3-vectors and for one dimensional space-time diagrams. Each point on a space-time diagram represents an event.

Space-time diagrams are defined such that lines of equal time are horizontal and lines of equal distance are vertical (figure 1). By definition, uniform motion in the reference frame is shown by a straight line on the diagram. To use radar we must know the speed of light (if distance were defined using a ruler, then to measure the time at an event we would still need to know the speed of a message from the event). But now we have a paradox. To measure speed we conduct a time trial over a measured distance, but first time must be defined at both ends of the ruler, which requires knowledge of the speed of light. We know no other way to measure the time of an event at a distance from a clock; if we synchronise two clocks by bringing them together, we have no guarantee that they remain synchronised when they are separated, unless light is used to test their synchronisation. Thus the speed of light is an absolute constant because measurement of 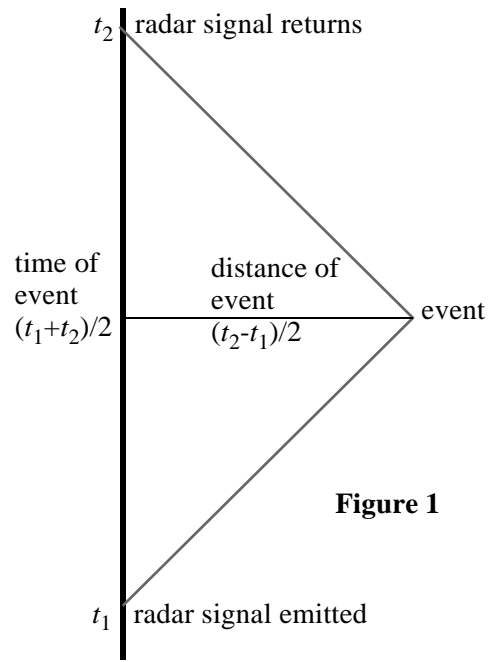

Figure 1

speed requires a co-ordinate system, which requires light for its definition. An experiment to determine the speed of light actually measures the conversion factor from natural units in which the speed of light is 1. By definition, light is drawn at 45° in a space-time diagram.

**Definition:** A space time co-ordinate system defined by radar is known as a reference frame.

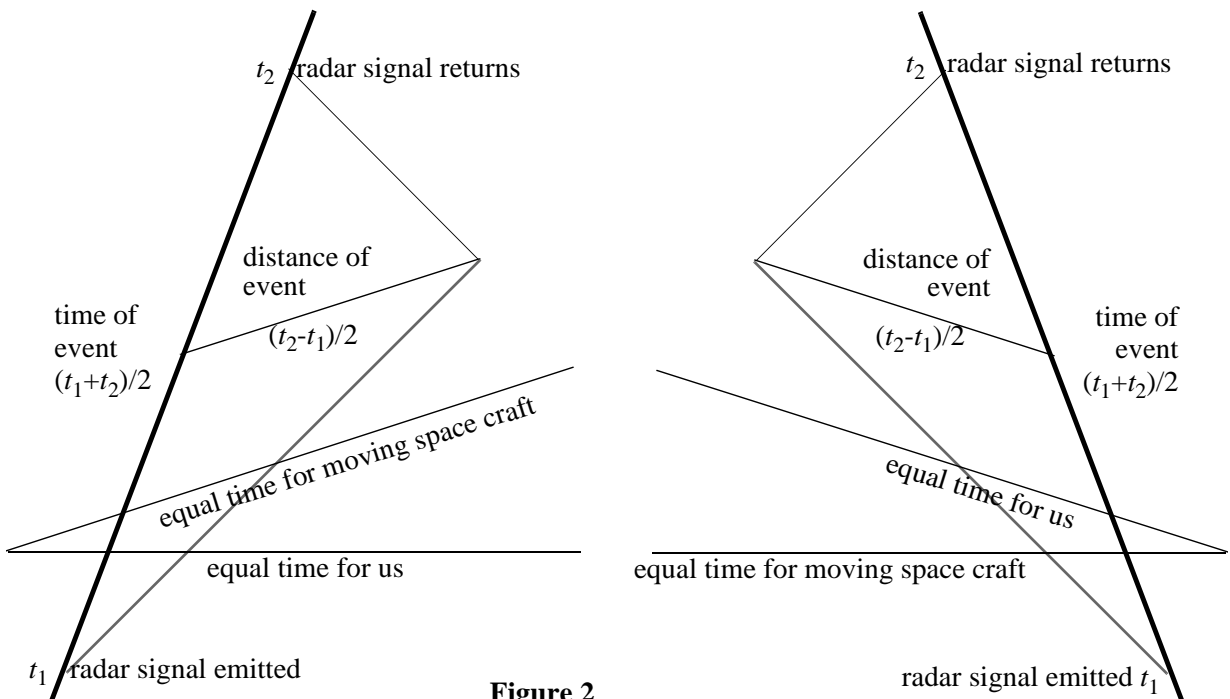

Figure 2

Once clocks are separated, there is no way to synchronise them directly, but, according the principle of homogeneity, two clocks will give the same unit of time if the physical processes in each are identical.

If we wish to compare our coordinate system with the coordinate system of a moving observer, we need to know what unit of time the moving observer is using. Once clocks are separated, there is no way to synchronise them directly, but, according the principle of homogeneity, two clocks will give the same unit of time if the physical processes in each are identical. Figure 2 shows the coordinate system defined by an observer in a moving space craft, as it appears to us, and our coordinate system as it appears to him. The moving observer represents himself with a vertical axis, and he would draw us at an angle. In his diagram our reference frame appears distorted.

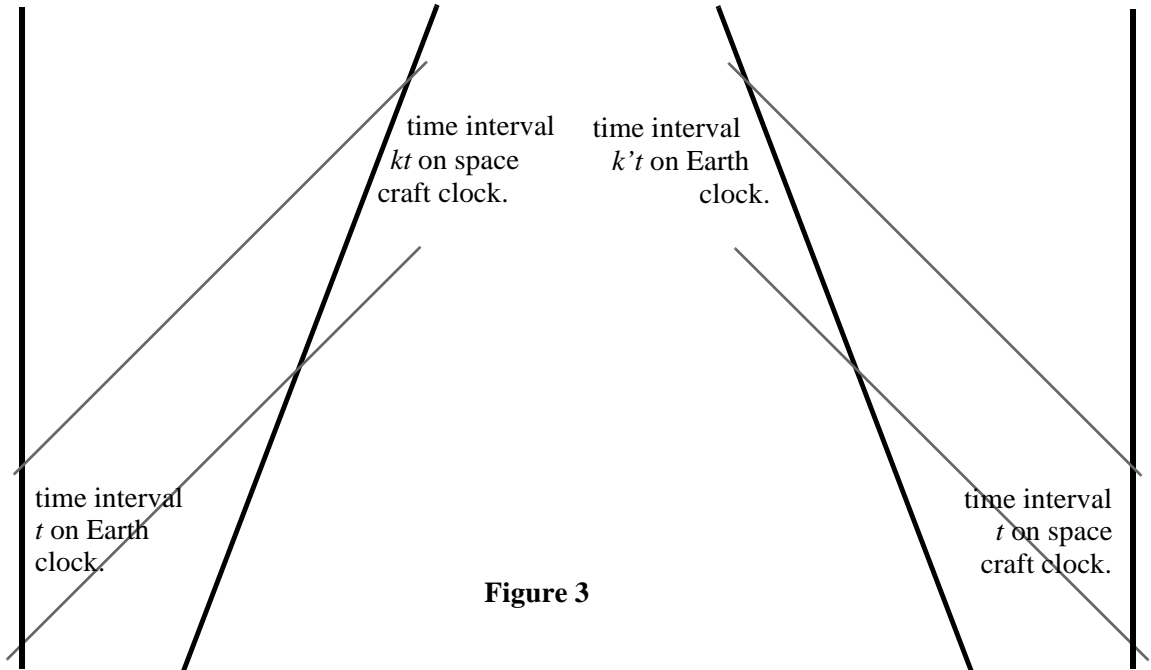

**Figure 3**

In figure 3, a space craft is uniformly moving in the Earth's reference frame. The space craft and the Earth have identical clocks and communicate with each other by radio or light. The Earth sends the space craft two signals at an interval $t$. The space craft receives them at an interval $kt$ on the space craft's clock. $k \in \mathbb{R}$ is immediately recognisable as red shift (by considering the signals as the start and stop of a burst of light of a set number of wavelengths of a set frequency). Similarly if the observer on the space craft sends two signals at an interval $t$ on his clock, they are received at an interval $k't$ on the Earth.

There is no fundamental difference between the matter in the space craft and the matter in the Earth. The space craft can be regarded as stationary, and the Earth as moving. The principle of homogeneity implies that signals sent by the space craft to the Earth are also subject to red shift. The defining condition for the special theory of relativity is that there is a special class of reference frames such that

**Definition:** For inertial reference frames red shift is both constant and equal for both observers, $k = k'$.

**Definition:** The law of co-ordinate transformation between inertial reference frames is Lorentz transformation.

We know from observation that inertial reference frames exist, at least to the accuracy of measurement and they will be assumed in this paper. The general theory of relativity places a more general condition on red shift. The implication is studied below in section 3, *Non-Euclidean Geometry*, where it is shown that an inherent delay in the return of the signal forces the use of non-inertial frames (such that $k \neq k'$) and results in the force of gravity.

**Theorem:** (Time dilation, figure 4) The time $T$ measured by a space craft's clock during an interval $t$ on the Earths clock is given by

2.1 $\qquad T = t\sqrt{1-v^2}$

**Proof:** The space craft and the Earth set both clocks to zero at the moment the space craft passes the Earth. The space craft is moving at speed $v$, so by definition, after time $t$ on the Earth clock, the space craft has travelled distance $vt$. Therefore Earth's signal was sent at time $t - vt$, and returned at time $t + vt$. For inertial reference frames, if the space craft sends the Earth signals at an interval $t$ the Earth receives them at an interval $kt$. So

2.2 $\qquad T = k(t - vt)$.

Then by applying the Doppler shift again for the signal coming back

2.3 $\qquad t + vt = k^2(t - vt)$

Eliminating $k$ gives 2.1, the formula for time dilation.

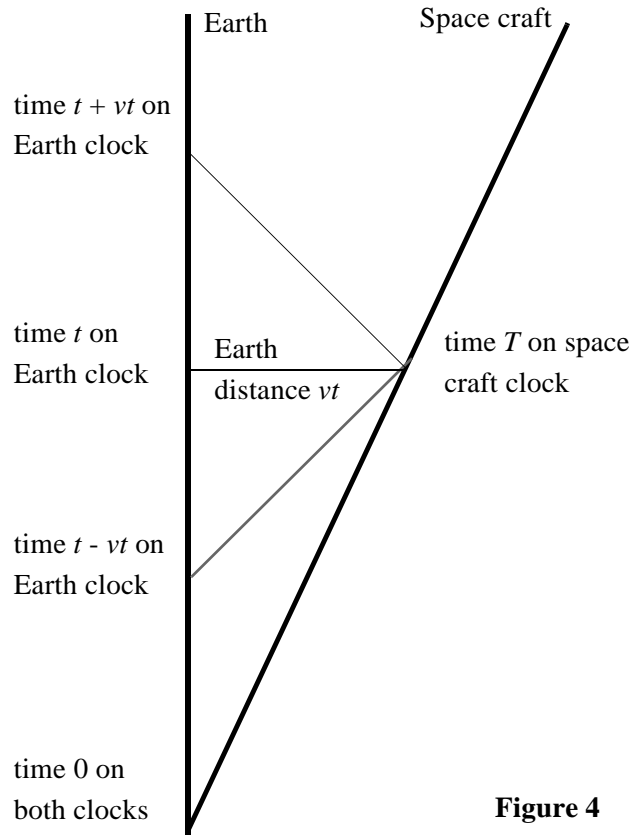

Figure 4

**Theorem:** (Lorentz Contraction, figure 5) A distance $d$ on the earth is measured on a space craft to be

2.4 $\qquad D = \dfrac{d}{\sqrt{1-v^2}}$

**Proof:** The bow and stern of the space craft are shown as parallel lines. The space craft's clock is in the bow. For ease of calculation, both the space craft and the Earth set their clocks to zero when the bow passes the Earth clock. Earth uses radar to measure the distance, $d$, to the stern at time 0. To do so, the signal must have been sent at time $-d$, and must return at time $d$ on the Earth clock. From the Doppler shift, on the space craft's clock, the signal passes the bow of the space craft at time $-d/k$ and comes back at time $dk$. So, according to the moving space craft

2.5 $\qquad D = (dk + d/k)/2$

Eliminating $k$ using 2.3 gives 2.4, the formula for Lorentz contraction.

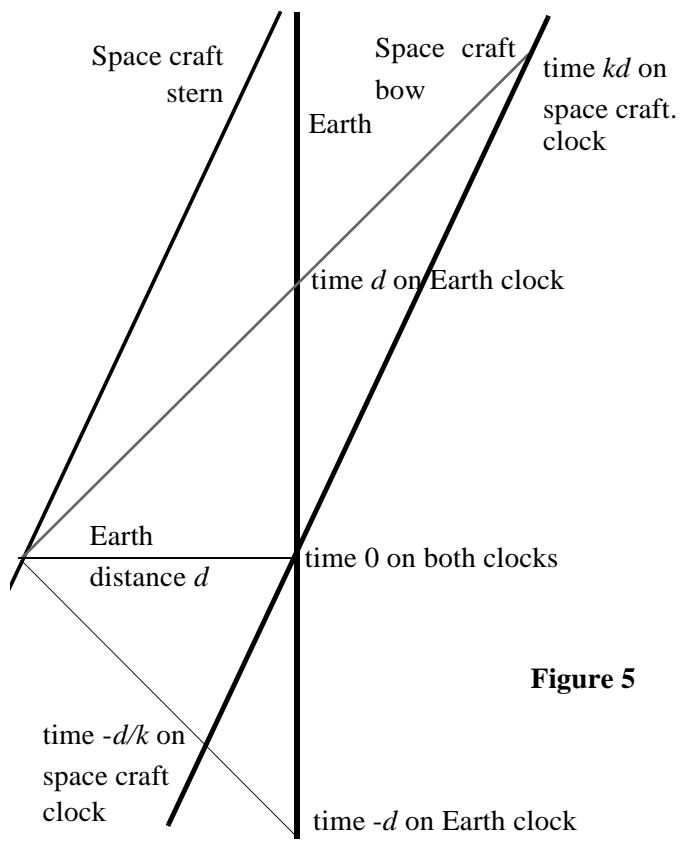

Figure 5

Laws which are the same in all co-ordinate systems are expressed in terms of invariants, mathematical quantities which are the same in all co-ordinate systems. The simplest invariant is an ordinary number or scalar. Another invariant, familiar from classical mechanics, is the vector. Changing the co-ordinate system has no effect on a vector, but it changes the description of a vector in a co-ordinate system.

**Definition:** A space-time vector is the difference in the co-ordinates of two events. When no ambiguity arises space-time vectors are simply called vectors.

**Theorem:** The mass shell condition

2.6 $$m^2 = E^2 - p^2$$

**Proof:** A vector can be represented as a straight line on a space-time diagram, and described by components

2.7 $$r = (E, p)$$

For a time-like vector, $r$, there is a particular reference frame in which it represents a state of rest, namely when it is aligns with the axis representing the clock on which the definition of that reference frame is based. In this reference frame $r$ has co-ordinates

2.8 $$r = (m, 0)$$

An observer moving at velocity $v$ relative to the clock describes $r$ by co-ordinates given by the formulae for time dilation, 2.1 and Fitzgerald contraction, 2.4

2.9 $$r = (E, p) = \left( \frac{m}{\sqrt{1-v^2}}, \frac{mv}{\sqrt{1-v^2}} \right)$$

The mass shell condition, 2.6, follows at once

**Definition:** If $x = (x_0, x_1, x_2, x_3)$ and $y = (y_0, y_1, y_2, y_3)$ are vectors in space-time then the scalar product is

2.10 $$x \cdot y = -x_0 y_0 + x_1 y_1 + x_2 y_2 + x_3 y_3$$

**Theorem:** The scalar product is invariant under Lorentz transformation

**Proof:** Straightforward algebra from 2.9

## 3 Non-Euclidean Geometry

When Euclid wrote down the axioms of geometry he was aware that neither he nor his contemporaries had been able to prove his fifth postulate, (that parallel lines can extend indefinitely always the same distance apart), but for two thousand years it was subjected to attempted proof. Newtonian space is rooted in Euclidean geometry, and as a metaphysical notion has been repeatedly and severely challenged by philosophers and mathematicians. In particular Gauss expressed severe doubts about the a priori truth of Euclidean geometry and was perhaps the first to believe that it was not necessarily true [2]. He organised expeditions into the Alps to measure whether the sum of the angles of a large triangle is $180^o$, but did not detect a deviation from Euclidean geometry.

For historical reasons non-Euclidean geometries are called curved. But this is not curvature in the familiar sense of a curved surface. We distinguish between internal and external curvature. External curvature is the familiar concept of curvature, the shape of a surface observed in three dimensional space. Internal or Gaussian curvature refers the the geometrical properties of a space. We will see that the curvature of space-time is internal, because it refers to geometrical properties defined by measurement within space-time. External curvature is not sensible when applied to space-time, because there is no out-side to look at it from. In general, non-Euclidean geometries are rigorously described by Riemann's theory of differential manifolds and the tensor calculus. But in this treatment we will use an intuitive characterisation based on red shift, which is adequate for many physical treatments.

If the geometrical properties of matter are conceived as empirical properties of particle interactions then there is no prior reason to believe in Euclid's fifth postulate. If we measure the distance AB between any two points, and measure two equal distances AD = BC = $h$ perpendicular to AB, as in figure 6, then we have no prior reason to assume that AB is equal to DC, as measured by an observer at D. Of course there is nothing to stop us from constructing a Euclidean co-ordinate system, in which the distance DC is defined to be the same as AB. It is often convenient to do so, but we do not assume that distances calculated in a Euclidean co-ordinate system are the same as distances found by direct measurement. We also cannot assume that it is meaningful to extend a Euclidean co-ordinate system indefinitely in all directions out into space; coordinates are only meaningful in regions of the universe where it is possible to (directly or indirectly) measure distances.

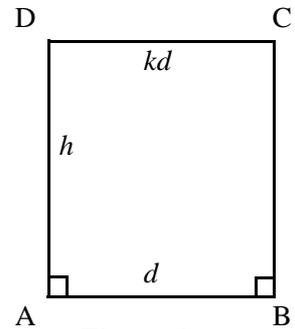

**Figure 6**

The Gaussian curvature of a geometry can be characterised by its geometric properties. For example we can characterise curvature by comparison with the parallel postulate: Let AD = $d$, and let CD = $kd$.

> Negative curvature:   $k$ increasing with $h$
>
> Zero curvature:   $k$ constant as $h$ increases
>
> Positive curvature:   $k$ decreasing as $h$ increases

The parallel postulate is a not an ideal criterion on which to base a categorisation of geometry, because geometrical systems start from a point, namely the origin of the co-ordate system, not a line. An equivalent, and more natural, characterisation of a geometry is found by considering the length of an arc, CD, of a circle of radius, $s$, subtended by an angle, $\theta$, at the origin, O (figure 7). Use a very small angle, $\theta$, and drop perpendiculars of equal length from CD to a base line, AB, through the origin O. Then, in Euclidean geometry the length of CD is $s\theta$, almost equal to AB. But in general the length of CD is $ks\theta$, and the value of $k$ characterises the geometry according to the relationships

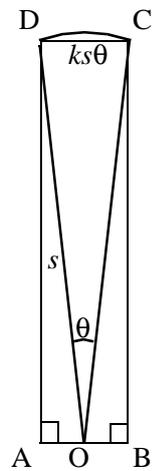

**Figure 7**

> Negative curvature:   $k$ increasing with $s$
>
> Zero curvature:   $k$ constant as $s$ increases
>
> Positive curvature:   $k$ decreasing as $s$ increases

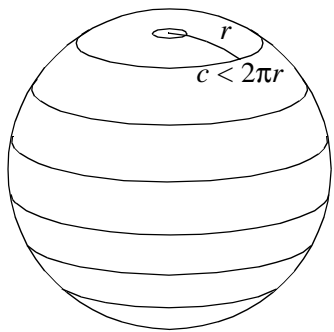

**Figure 8**

The sphere has positive curvature, since the ratio $c/r$ decreases as $r$ increases (figure 8). It is locally flat; $c/r \to 2\pi$ as $r \to 0$. The cone is flat everywhere except at the apex, where the geometry has a cusp, or singularity (figure 9). A circle enclosing the apex has circumference less than $2\pi r$, a circle anywhere else has circumference equal to $2\pi r$.

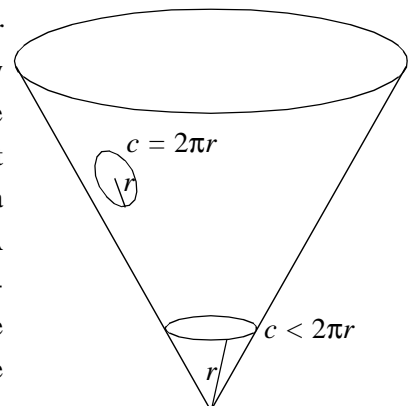

**Figure 9**

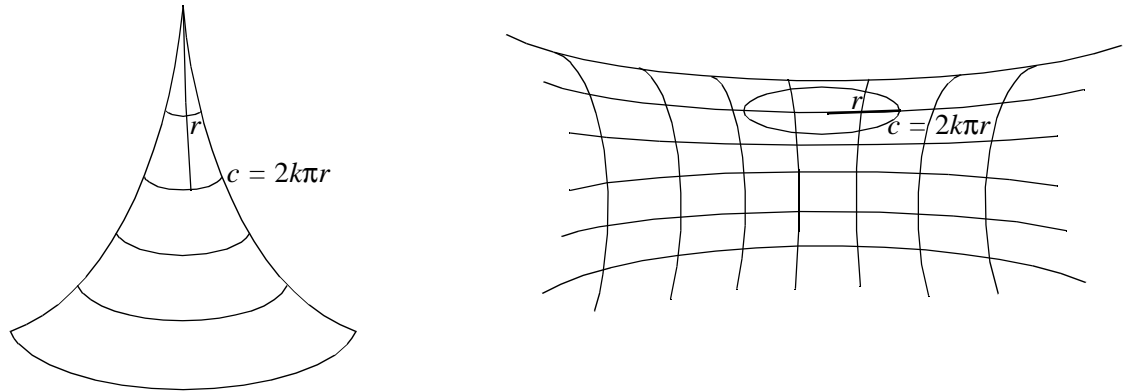

**Figure 10**  Trumpets and saddles have negative curvature; $k$ increases with $r$.

I

In the geometry of space-time, distances are defined in terms of time on a given clock. We cannot assume that two identical clocks will keep time when separated. By drawing a space-time diagram (figure 11), again defined so that light is drawn at $45^o$, we see that the change in speed of the clock is determined by red shift, $k$ (by considering the signals as the start and stop of a burst of light of a set number of wavelengths of a set frequency). Then distances measured by radar from a given clock are not necessarily the same as those measured from another clock. To find a definition of distance with which to characterise the geometry, we define the metric distance as the distance along a given path where each short length along the path is measured from a clock on that position of the path. It should be obvious that some form of limit is involved in this definition of distance, and in truth the proper expression of the metric distance in Riemann's geometry is not easy, but for this paper we will simply accept an intuitive notion of the limit.

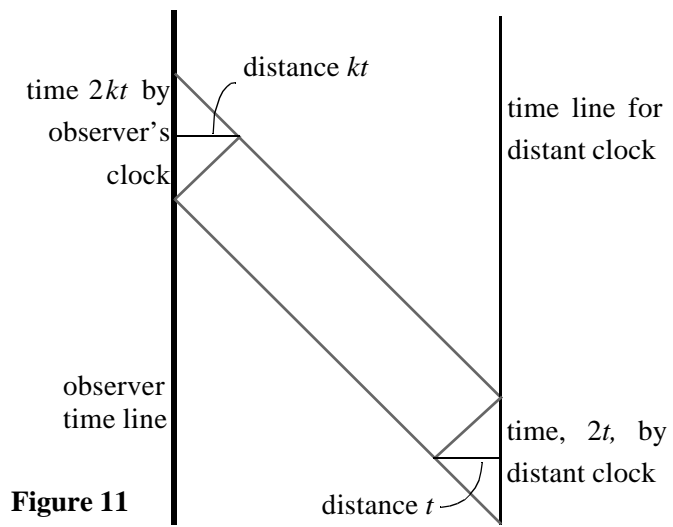

**Figure 11**

## 4  Space-time Geometry

To imagine a substructure for space-time we conceive that each elementary charged particle follows some repeating process according to which we may regard a primitive notion of time as one of its fundamental properties. We call this notion a time line. For example a particle may have the possibility of emitting or absorbing a photon in each discrete instant of its time line. This can be considered a repeating process adequate for the notion of primitive time. Whenever an exchange of photons takes place, i.e. a photon is emitted by one charged particle and absorbed by another, and a photon is then immediately emitted by the second particle and absorbed by the first, then a coordinate for the second particle is established in terms of the time line of the first.

Since visualisation involves the awareness of geometry, it is clear that the pre-geometric properties of matter cannot strictly be visualised. Nonetheless, a "stitch in space-time" can be illustrated diagrammatically (figure 12). Charged particles are shown as dashed lines, where the dashes represent discrete intervals of time in the particle's time line. Photon exchange is shown by continuous grey lines. A single stitch such as that shown can only give a single value of distance and time for the second particle. Many stitches are required to determine the properties of space-time. It is not necessary to assume they always require the immediate return of a photon, only that they use photon exchange and combine to a consistent geometry.

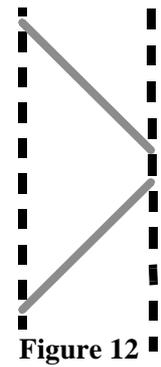
Figure 12

It is legitimate to imagine a system of such particles in the form of a diagram provided that it is understood that the space between the lines of the diagram has no practical meaning. The properties of space-time depend on internal relationships between dashes and nodes in the diagram, not on the geometry of a drawing. Thus in figure 13 the increased lapsed time for the return of a photon indicates that the particles are moving away from each other. In systems of many particles such as we generally observe, photons are constantly exchanged, and macroscopic space-time can be construed as some kind of composition or average of the primitive space-time associated with photon exchange. Since the process of photon exchange is the same as we use in radar, the average behaviour of a system in which there are many such exchanges should obey the geometrical relationships found by means of radar.

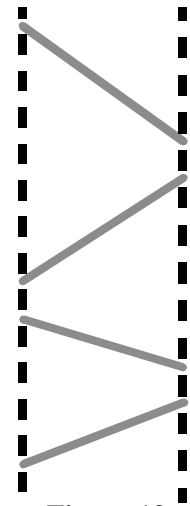
Figure 13

It is only necessary to calculate the gravity due to the curvature in space-time caused by a single elementary particle. It will be seen that the geometry of space-time has a singularity of total curvature proportional to mass-energy at each elementary particle, and is Lorentzian elsewhere. In a macroscopic reference we measure averaged effects, so this appears as curvature, not as a singularity. The linearity of tensors is then sufficient to state Einstein's field equation for general relativity [6] (without the cosmological term)

A photon is emitted by **A**, and absorbed by a particle **B** of mass $m$, and one is reflected back to **A** (figure 14). There is an inherent delay in the reflection due to the discrete nature of particle interactions. It is not obvious that there is a single interval of time between the absorption and emission of the photon but there must be a characteristic lag of magnitude $T$ for the reflection to take place (the inherent lag in return of radar is a feature of the interaction between elementary particles, and applies to macroscopic phenomena only as the expectation of many elementary particle interactions). The lag can be written as a vector in **B**'s reference frame, and is therefore proportional to the energy momentum vector in that frame. Thus, for some constant $G$

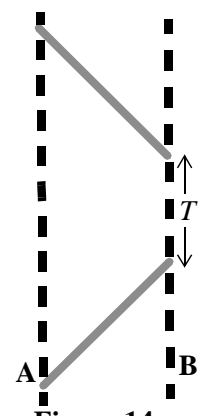
Figure 14

4.1 $\qquad T = (T, \mathbf{0}) = G(m, \mathbf{0})$

After Lorentz transformation this appears in **A**'s reference frame as

4.2 $\qquad T = G(M, \mathbf{P})$

where $M$ is the mass-energy and $\mathbf{P}$ is the momentum of **B**.

If space-time diagrams were an accurate representation, then the co-ordinate system could be perfectly mapped onto the Euclidean geometry of a flat piece of paper. But the reflection of light is really two events, the absorption and the emission of a photon (figure 15). These two events have the same space-time coordinate. We cannot redefine geometry by simply subtracting an additional distance *GM* to the distance calculated from radar for three reasons. First, the value of time would still be poorly defined at the event. Second, we have no knowledge of what the value of *GM* actually is; it may have a different value depending on the type of particle that the photon reflects off. Third, subtracting *GM* does not permit a consistent definition of geometry; before we subtract *GM* we must know what red shift to apply, and before we know red shift we must have defined geometry.

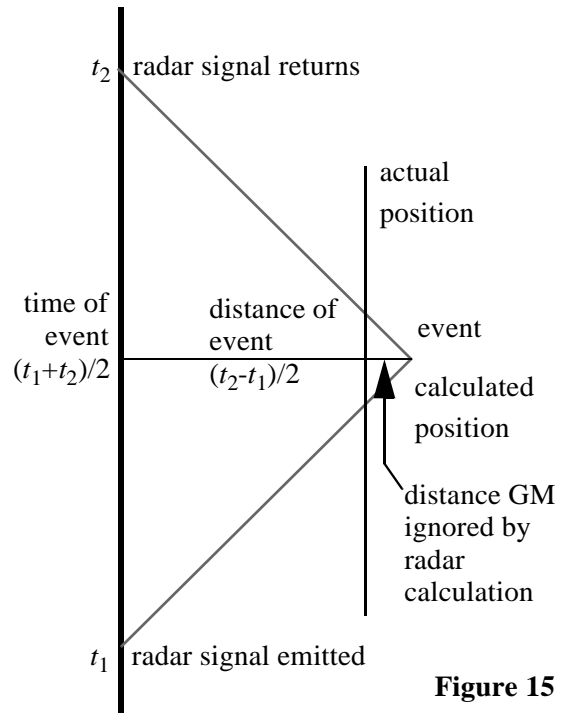

**Figure 15**

We can calculate the effect on the geometry of space-time by comparing the distance, *r*, of an event as measured by radar, with the metric distance, *s*, measured along a path from the event to the clock and calculated by adding together small distances along the path, each measured individually by radar in that part of the path. Figure 16 is a space-time diagram scaled so that light is always drawn at 45°. Then an observer at a distance from an event sees a shift $k$ in frequency of a clock at the event, shown by the time intervals, *s* and $r = ks$, measured by each clock between the signals from the event to the observer. For $k < 1$ (red shift) this can be understood as the observer's clock going fast. To determine curvature we calculate $k$.

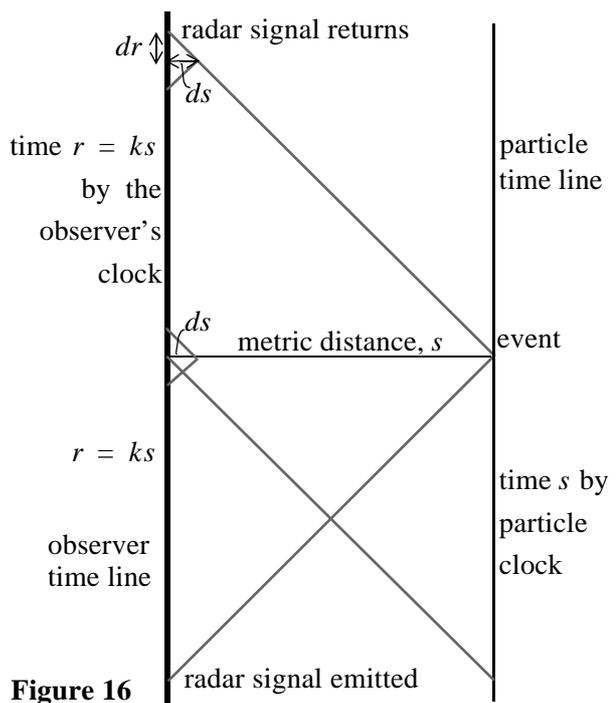

**Figure 16**

By definition.

4.3    $r = ks$

Differentiating,

4.4    $dr = sdk + kds$

Because each small distance, *ds*, is calculated by direct measurement (i.e. radar, shown at the top of figure 12.) we have

4.5    $ds = dr$

Substituting and rearranging the terms,

4.6 $$\frac{1}{s}ds = \frac{1}{1-k}dk$$

Integrating

4.7 $$\ln s = \ln\left(\frac{A}{1-k}\right)$$

where $A$ is a constant to be determined. So

4.8 $$s = \frac{A}{1-k}$$

and by 4.3 we obtain

4.9 $$r = s - A$$

The minimum distance which could be measured by radar is $r = GM$, i.e. the characteristic lag at $s = 0$. Thus $A = -GM$ and we have

4.10 $$r = \left(1 + \frac{GM}{s}\right)s$$

Comparison with 4.3 gives the formula for red shift

4.11 $$k = \left(1 + \frac{GM}{s}\right)$$

Apply this red shift to the wave function for every particle in an isolated body of initial energy $E_0$, and take the expectation to find the classical energy equation for a body in a spherical gravitational field.

4.12 $$E = E_0\left(1 + \frac{GM}{s}\right)$$

By summing the red shifts generated by every elementary particle in the universe we find the law of universal gravitation.